\documentstyle[prl,aps]{revtex}

\def\be {\begin{equation}}
\def\ee {\end{equation}  }
\def\beq{\begin{eqnarray}}
\def\eeq{\end{eqnarray}  }
\def\bi {\begin{itemize} }

\def\ei {\end{itemize}   }
\def\RE {I\kern-6pt R    }
\def\Z  {Z\kern-13pt Z   }
\def\be {\begin{equation}}
\def\ee {\end{equation}  }
\def\beq{\begin{eqnarray}}
\def\eeq{\end{eqnarray}  }

\def\eeq{\end{eqnarray}  }

\input epsf

\begin{document}
\draft

\twocolumn[\hsize\textwidth\columnwidth\hsize\csname
@twocolumnfalse\endcsname

\title{Static Gravitational Global Monopoles}
\author{Steven L. Liebling}
\address{Theoretical and Computational Studies Group\\
     Southampton College, Long Island University,
     Southampton, NY 11968}

\maketitle

\begin{abstract}
Static solutions in spherical symmetry are
found for gravitating global monopoles.
Regular solutions  lacking a horizon are found for $\eta < 1/\sqrt{8\pi}$,
where $\eta$ is the scale of symmetry breaking.
Apparently regular solutions with a horizon are found
for $1/\sqrt{8\pi} \le \eta \alt \sqrt{3/8\pi}$.
Though they have a horizon, they are not Schwarzschild.
The solution for $\eta = 1/\sqrt{8\pi}$ is argued to have
a horizon at infinity.
The failure to find static
solutions for $\eta  > \sqrt{3/8\pi} \approx 0.3455$ is consistent with
findings that topological
inflation begins at $\eta \approx 0.33$.
\end{abstract}


\vskip2pc]

Topological defects have attracted quite a bit of attention
because of their relevance to a number of different areas 
ranging from condensed matter to structure formation.
Studies of global monopoles in particular have served as
a foundation on which knowledge of other defects 
has been built.
Previous work
details static gravitating global monopole solutions~\cite{barriola,lousto}, while
further studies
consider the
gauged case~\cite{maison1,ortiz,lee}. Understanding of the static
solutions is also relevant
to the study of topological inflation~\cite{alex,linde}.

Here I return to the global monopole case and consider the possibility
that static global monopoles have a horizon
(the ``rather curious'' monopoles mentioned in~\cite{lousto}).
I reproduce the solutions of~\cite{barriola}
for $\eta < 1/\sqrt{8\pi}$, and find other solutions for 
$1/\sqrt{8\pi} \le \eta \alt \sqrt{3/8\pi}$ which, though they contain
a horizon, appear regular. I also comment on the possibility that
the failure to find
static solutions for $\eta > \sqrt{3/8\pi}$ is indicative of the
onset of topological inflation which has been reported
for $\eta \agt 0.33$~\cite{cho,sakai,sakai3}.

Letting $\Phi^a$ represent a triplet scalar field and including
the usual symmetry breaking potential with scale of symmetry breaking $\eta$,
the Lagrangian is
\be
L =
     - \frac{1}{2} \Phi^a{}^{;\mu} \Phi^a{}_{;\mu}
     - \frac{1}{4} \lambda \left[ \left( \Phi^a \right)^2 - \eta^2 \right]^2,
\label{lagrangian}
\ee
where $\lambda$ is a coupling constant which sets the scale.
Henceforth, I choose $\lambda=0.1$ without loss of generality.

\begin{figure}
\epsfxsize=8cm
\centerline{\epsffile{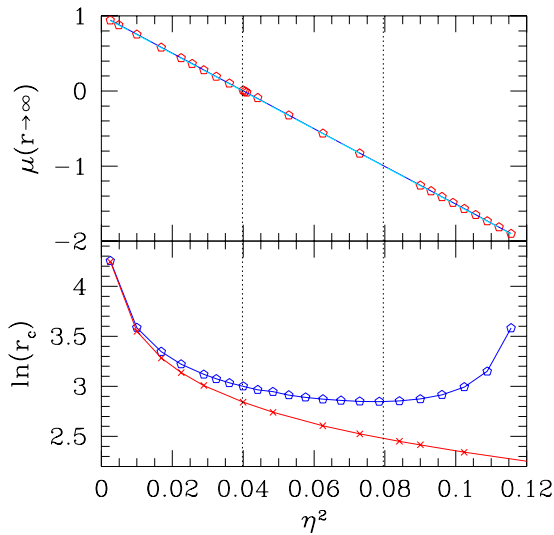}}
\caption{The top frame shows the asymptotic behavior
         of $\mu(r)$.
         The points represent the 
         asymptotic value of $\mu$ computed via the slope of a least-squares
         fit to $r\mu$ versus $r$ for each $\eta$.
         The solid line shows the least-squares fit to the points,
         while the dashed line (indistinguishable from the fit)
         shows the expected relationship
         $\mu(r\rightarrow\infty) = 1 - 8\pi \eta^2$.
         The vertical dotted lines denote  $\eta^*$
         and $\eta^\natural$.
         The bottom frame shows the core radius (open pentagons)
         which reaches a minimum at $(\eta^\natural)^2 \approx 0.08$.
         For comparison, the core radius for flat-space
         monopoles (cross hatches)
         is also shown.}
\label{fig:asymp}
\end{figure}

The spherically symmetric metric
\begin{equation}
ds^2 = - A^2 \mu~dt^2
       + \frac{1}{\mu}~dr^2
       + r^2~d\Omega^2
\label{eq:metric}
\end{equation}
is adopted in terms of the metric components $A(r)$ and $\mu(r)$
(the same as in~\cite{maison1} modulo the sign convention).
By association with the Schwarzschild  metric, a mass aspect function
$m(r)$ is defined
\begin{equation}
m(r) \equiv \frac{r}{2} \left( 1 - \mu \right).
\end{equation}
The usual hedgehog ansatz for the triplet field, $\Phi^a=f(r)~\hat r$,
is chosen in terms
of the monopole profile $f(r)$.
Casting the equations in first-order form,
an auxiliary variable $\Psi(r) \equiv f'$ is introduced
(a prime
denotes $d/dr$).
The equations for static solutions then become
\beq
f'      & = &  \Psi \\
\Psi'   & = &
              \frac{f}{r^2 \mu} \left[
                                       2
                                      +\lambda r^2 \left( f^2 - \eta^2 \right)
                                        \right]
              -\Psi \left(
                            \frac{2}{r}
                          + 4\pi r \Psi^2
                          + \frac{\mu'}{\mu}
                          \right)
\\
\mu' & = &
            \frac{1-\mu}{r}
          - 4\pi r \Psi^2 \mu
          - 8\pi \left[
                           \frac{f^2}{r}
                         + \frac{\lambda r}{4} \left(
                                                     f^2 - \eta^2
                                                     \right)^2
                        \right]
\label{eq:mu}
\\
A' & = & 4 \pi r A \Psi^2.
\eeq
Imposing regularity as $r\rightarrow\infty$, these equations show
that $\mu'\rightarrow 0$ and $\mu \rightarrow 1-8\pi \eta^2$.
Confirmation that the numerical solutions obey this behavior is
shown in Fig.~\ref{fig:asymp}.
Thus, the space has a deficit solid angle
$\Delta = 4\pi \left( 8\pi \eta^2 \right)$~\cite{cho}.

\begin{figure}
\epsfxsize=7cm
\centerline{\epsffile{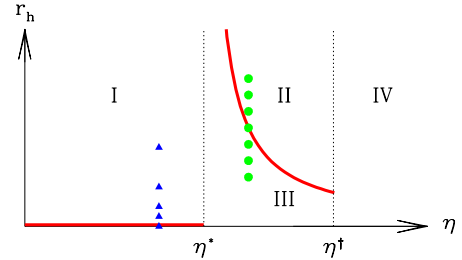}}
\vspace{-1.05in}
\caption{Schematic diagram of the solution space of
         static solutions. Regular solutions are represented
         by the two disjoint bold curves.  The radius of any horizon $r_h$
         is plotted versus $\eta$.
         The regions denote the types of singular solutions:
         (I) black hole solutions, (II) a single horizon, (III) two
         horizons, (IV) no static solutions found.
         The family of solutions denoted by solid
         triangles is shown in Fig.~\ref{fig:eta=0.15}. The family
         denoted by solid circles is shown in Fig.~\ref{fig:eta=0.25}.
         }
\label{fig:schematic}
\end{figure}

These equations have singularities at $r=0$ and where $\mu=0$.
To integrate outward from $r=0$, regularity is assumed, and by
Taylor expanding about $r=0$, the solutions can be integrated 
from close to the origin. Specifically, the conditions
\beq
\mu(0) = 1 & ~~~~ & \mu'(0) = 0 \\
f(0)   = 0 & ~~~~ & \Psi'(0) = 0
\eeq
apply, and $\Psi(0)$ is a free parameter which is adjusted
via a standard shooting method
until the correct asymptotic behavior for $f(r)$ is observed,
specifically $f(r\rightarrow \infty)=\eta$.

Solutions where $\mu(r)$ vanishes can be handled in a similar
fashion. Defining $r_h$ to be the radius of the horizon such
that $\mu(r_h) \equiv 0$, appropriate boundary conditions at $r=r_h$
can be found by enforcing regularity there.
Then by Taylor expanding about $r_h$, the solutions can be integrated
either outward or inward from
near the horizon. The value $f(r_h)$ is then a
free parameter which is adjusted via a shooting method
so that, if integrating outward, the solution 
satisfies $f(r\rightarrow \infty)=\eta$, or,
if inward, satisfies $f(r\rightarrow 0) = 0$.
A standard ODE integrator has been used.

The solutions found are summarized in a schematic of the solution
space in Fig.~\ref{fig:schematic}. Regular solutions lacking a horizon
are found as $\eta$ is increased to a critical value,
$\eta^* \equiv 1/\sqrt{8\pi} \approx 0.1995$.
For $\eta \ge \eta^*$, static solutions
with an apparently regular horizon are found up to $\eta\approx 0.3455$.
Empirically it  appears that
this upper limit for $\eta$ occurs at $\sqrt{3/8\pi}$,
although unlike the case of $\eta^*$ no theoretical justification for this
limit is found.
Above this second critical value $\eta^\dagger \equiv \sqrt{3/8\pi}$,
no static solutions are found. In addition to the
regular static solutions, singular solutions are also found as discussed
below.

\begin{figure}
\epsfxsize=8cm
\centerline{\epsffile{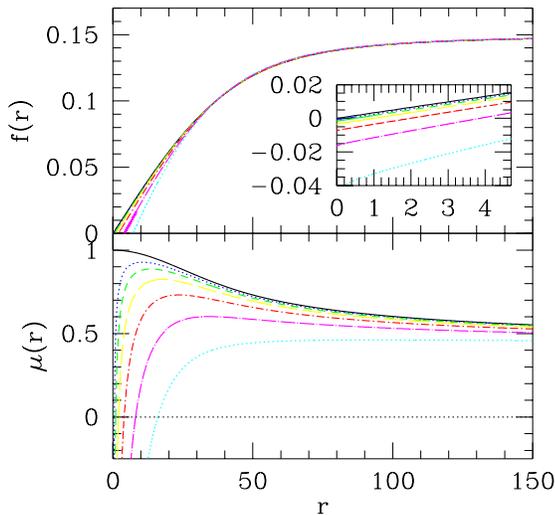}}
\caption{Family of static solutions for $\eta=0.15$.
         One solution (solid) is found to be regular
         and static, while the other solutions contain
         a horizon at $r_h = 0.5, 1, 2, 4, 8, 16$
         and are singular at the origin.
         Where $\mu(r)$ vanishes denotes a horizon.
         }
\label{fig:eta=0.15}
\end{figure}

For $\eta < 1/\sqrt{8\pi}$, horizonless static solutions exist
as described in~\cite{barriola}. A typical example
of such a solution is shown in Fig.~\ref{fig:eta=0.15}.
Other solutions are also shown which are singular at $r=0$ and correspond to
black holes containing a monopole charge. These singular solutions
are obtained by enforcing the existence of a horizon at a particular
radius and demanding that $f(r\rightarrow \infty)=\eta$.
For a given value of $\eta$, a family of static solutions exists,
only one member of which is regular.

Fig.~\ref{fig:eta=0.15}  shows that asymptotically the solutions approach
one another independent of the existence of the black hole.
Such a convergence of the singular solutions to the regular solution
for $r_h \rightarrow 0$ is also observed for the gauged monopole
in~\cite{maison1}.
Furthermore, in contrast to the solution for a gauged
monopole, the metric component $\mu$
does not asymptote to unity as it would in an asymptotically flat
spacetime. Instead,
that it asymptotes to a non-unit value indicates
the linear
divergence of the mass of an isolated global monopole.

Solutions similar to those shown in Fig.~\ref{fig:eta=0.15}
are found for $\eta$ increasing until $\eta \approx 0.20 \approx \eta^*$.
As $\eta$ is increased, the asymptotic value of $\mu$ decreases
toward zero. For $\eta \agt 0.20$, no solutions without $\mu$
vanishing are found. This result agrees with the argument
presented in~\cite{cho} that when $\eta > 1/\sqrt{8\pi}$ no static
(horizonless) solutions exist.

For the critical case in which $\eta=\eta^*$,
the solution 
has $\mu$ vanishing at $r=\infty$. 
This solution then represents a static spacetime with a horizon at
infinity.
It is not clear if there are any cosmological implications
to the existence of such a solution.

\begin{figure}
\epsfxsize=8cm
\centerline{\epsffile{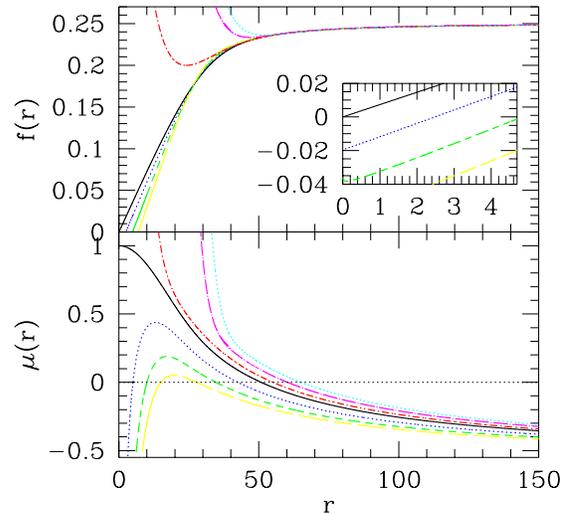}}
\caption{Family of static solutions for $\eta=0.25$.
         One solution (solid) is found to be regular
         and has a horizon at $r_h=50.78$.
         Three solutions singular at the origin are shown
         containing two
         horizons, the first occurring at $r_h = 5, 10, 15$.
         Three other singular solutions
         having only one horizon
         are shown with $r_h = 55, 60, 65$.
         Compare to the sub-critical case shown in Fig.~\ref{fig:eta=0.15}.
         }
\label{fig:eta=0.25}
\end{figure}

For $\eta > \eta^*$, static solutions can be found, albeit with a horizon.
One such example is displayed in
Fig.~\ref{fig:eta=0.25}. Shooting from the origin, the radius
at which $\mu$ vanishes, $r_h$, can be determined. Then, by Taylor
expanding about $r_h$, the solution can be extended to large $r$.
The solution is thus regular both at $r=0$ and at $r=r_h$. Once
again, irregular solutions can be constructed by enforcing
the vanishing of $\mu$ at some other radius.
Solutions with horizons smaller than that of the regular solution
have two horizons as shown in the figure, while those with horizon
greater than that of the regular solution have only one. 

I note
that none of these solutions represents  a black hole as $\mu$
is negative for large $r$ and positive for small $r$, the opposite
of Schwarzschild. As is evident from the metric~(\ref{eq:metric}),
the roles of $t$ and $r$ switch outside the vanishing of $\mu$.
Hence, the exterior is no
longer static though the solution remains
independent of $t$. However, within the horizon these
solutions remain static.

\begin{figure}
\epsfxsize=8cm
\centerline{\epsffile{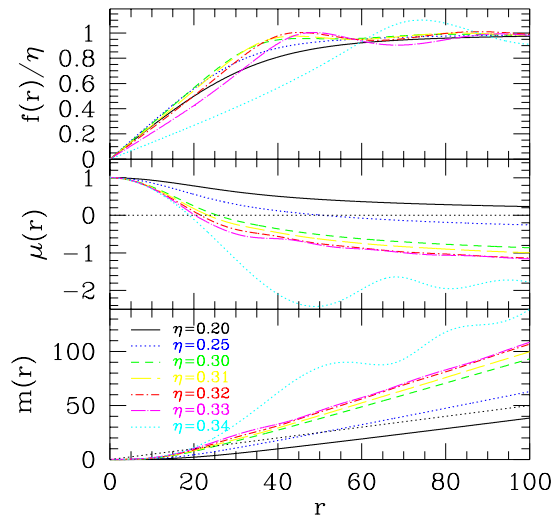}}
\caption{
         Set of static, regular solutions for $\eta\in [0.20,
         0.34]$.
         The top frame shows the rescaled hedgehog profile.
         The middle frame shows the metric component $\mu$
         with horizons indicated by $\mu=0$ (the dotted line).
         The bottom frame shows the mass aspect function
         with horizons indicated by $m=r/2$ (the dotted line).
         Note that the core radius decreases and then increases
         as $\eta$ increases, the transition occurring for $\eta\approx 0.28$.
         }
\label{fig:static_family}
\end{figure}

As $\eta$ is increased, another transition
is evident near $\eta\approx 0.28$. This transition occurs
at a new critical value of $\eta$, namely $\eta^\natural \equiv \sqrt{2/8\pi}$,
and is observed
by examining the size of the monopole.
Defining  the core radius
by $f(r_c) \equiv \eta/2$, I show
$r_c$ versus $\eta$ in Fig.~\ref{fig:asymp}. For
$\eta < \eta^\natural$,  the core radius decreases with increasing
$\eta$, while for $\eta > \eta^\natural$ the radius decreases.
Oscillations in  the solution
also become evident for $\eta > \eta^\natural$ as shown in
Fig.~\ref{fig:static_family}.

These oscillatory solutions bear a striking resemblance to the
reported stable solutions found for the dynamical evolutions
of gauged monopoles with large $\eta$
in~\cite{sakai}.
The oscillations become more pronounced as $\eta$
approaches $\eta^\dagger$, and is accompanied by
a dramatic decrease in the minimum to which $\mu$ reaches.
For $\eta > \eta^\dagger \approx 0.3455$
no static solutions are found.

\begin{figure}
\epsfxsize=8cm
\centerline{\epsffile{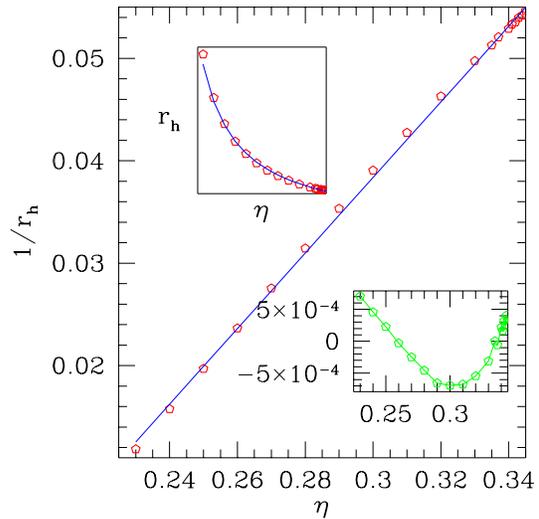}}
\caption{Behavior of the horizon radius $r_h$ of the static
         super-critical solutions versus $\eta$.
         The solid line indicates a least-squares fit of
         $1/r_h = A \eta + B$ where $A=0.3690$ and $B=-0.0723$.
         The upper insert shows the same data where
         $r_h \propto  1/(\eta - C)$ where $C=-B/A=0.1959$.
         It is expected that this value of $C$ would be $\eta^*
         \approx  0.1995$.
         The lower inset displays the
         deviation from the fit.}
\label{fig:r_h}
\end{figure}

Interestingly for super-critical solutions, the 
horizon $r_h$ of the regular solutions obeys a scaling law
in $\eta$. In particular, the horizon radius is found to obey
\begin{equation}
r_h \propto \frac{1}{ \eta - \eta^*}
\end{equation}
as demonstrated
in Fig.~\ref{fig:r_h}.

A summary of 
the solutions regular at the origin
are shown in
Fig.~\ref{fig:static_family}.
As $\eta$ is increased, the asymptotic value of $\mu$
is seen to decrease below zero indicating the presence of a horizon.
As $\eta$ is increased
further, the value of $\mu$ continues to decrease until 
$\eta = \eta^\dagger\approx 0.3455$ above which no static solutions are found.

The three critical values of $\eta$ all correspond to integer multiples of
a deficit solid angle of $4\pi$.
The deficit solid angle $\Delta^*$
occurring
for $\eta = \eta^*$ is known to be precisely $\Delta^* \equiv 4\pi$.
This critical value denotes the transition to static solutions with
horizons. The next transition occurs when the monopole core radius
changes from decreasing to increasing with $\eta$,
namely $\Delta^\natural \equiv 2\Delta^*$. Finally, the transition
above which no static solutions are found
occurs at $\Delta^\dagger \equiv 3 \Delta^*$. These latter two
transitions are found only empirically, calling for a geometric
explanation.

Having the static solutions in hand, the next question to consider
is whether they are stable. In particular, a relevant question
is whether the sub-critical static solutions are unstable to
collapse to a black hole. Ortiz asks this question of the
gauged monopoles, and answers it by considering the mass of the
various solutions~\cite{ortiz}. Where the monopole has greater mass than a
black hole with the same topological charge, the solution would
be expected to be unstable.

However, the mass of the global monopole is divergent,
and so it is not clear if such an argument can be made here.
The quantity $2m/r$, equal to $1-\mu$, does asymptote
to a finite value. Perhaps comparing the mass within a particular
radius would be sufficient to answer the question.
Fig.~\ref{fig:eta=0.15} shows, for
the sub-critical case, the behavior of $\mu$
for both the regular solution and various black hole solutions.
The black hole solutions do have more mass (smaller $\mu$) than
the regular solution for finite radius (i.e. ``locally''),
however they asymptote to the same value at infinity.
Independent of whether one looks at the local or asymptotic
value of $2m/r$, the regular solutions do not have greater mass than the
black hole, and this fact
might be some
indication that the solutions are indeed stable. Evolutions conducted
in~\cite{steve} also indicate that the solutions are stable.

Considering now the super-critical case, are these static solutions
unstable to some other solution? From Fig.~\ref{fig:eta=0.25},
one could consider the stability of the regular solution to the
irregular ones having either one or two horizons. However,
the physical significance of those horizons is not clear.

Instead, it is more interesting to examine these results in
the context of topological
inflation~\cite{alex,linde,cho,sakai,sakai3}.
As reported in~\cite{cho,sakai3}, when $\eta \agt 0.33$
the region inside a global monopole necessarily undergoes inflation.
The square of this value falls roughly in the middle of
the two transitions $\left(\eta^\natural\right)^2$ and 
$\left(\eta^\dagger\right)^2$, and thus it appears that topological
inflation does not begin at either of the transitions.

Instead, a change in stability appears
likely near $\eta\approx 0.33$.
A linear perturbation analysis should be able to confirm  both
the change in stability and the critical value of $\eta$
for which topological inflation begins.


{\em Acknowledgments:}
I am grateful for helpful discussions with Arvind Borde, Inyong Cho,
Eric Hirschmann, Nobuyuki Sakai, and Alexander Vilenkin.
I am also thankful for the support of the Southampton College Research
\& Awards Committee.


\end{document}